# AN X-RAY REGENERATIVE AMPLIFIER FREE-ELECTRON LASER WITH A STEP-TAPERED UNDULATOR

H.P. Freund,[1,2] P.J.M. van der Slot,[3] and P.G. O'Shea[1,4]

[1]Institute for Research in Electronics and Applied Physics, University of Maryland, College Park, MD 20742, USA
[2]NoVa Physical Science and Simulations, Vienna, Virginia 22182, USA
[3]Nonlinear Nanophotonics, MESA+ Institute, University of Twente, P.O. Box 217, 7500 AE Enschede, the Netherlands
[4]Department of Electrical Engineering, University of Maryland, College Park, Maryland 20742, USA

Regenerative amplifier x-ray free-electron lasers constitute alternative configurations for fourth-generation light sources which are typically operated in self-amplified spontaneous emission (SASE) mode. However, SASE exhibits relatively large fluctuations in the power and spectral properties from shot to shot, whereas XRAFELs provide a much more stable source of hard x-ray photons. Here, we study the performance of an x-ray cavity under consideration at the Stanford Linear Accelerator Facility using both uniform and step-tapered undulators with either transmissive or hole out-coupling.

The Linac Coherent Light Source (LCLS) at the Stanford Linear Accelerator Center (SLAC) [1] was the first operational x-ray Free-Electron Laser (XFEL), which produced 20 GW/83 fs pulses at a 1.5 Å wavelength with a repetition rate of 120 Hz operating in self-amplified spontaneous emission (SASE) mode. Since that time, numerous, mostly SASE, XFEL facilities have been constructed worldwide [2-9]. While SASE XFELs provide high-power sources of hard x-rays that enable a wide range of novel research opportunities, these light sources are subject to fluctuations in the output power and spectra of the order of 10 – 15% from shot to shot. As such, there is active interest in novel configurations to improve the stability of the output power and spectrum, spectral purity, and temporal characteristics with respect to SASE. Recently, it has been demonstrated that the longitudinal coherence can be improved in a SASE XFEL by the insertion of chicanes between the undulators [10]. Another possible way to improve longitudinal coherence is by means of a regenerative amplifier.

A regenerative amplifier (RAFEL) is a high-gain/low-$Q$ oscillator [11-20] that is expected to have improved temporal coherence over that of a SASE XFEL. This is particularly timely because an x-ray RAFEL (XRAFEL) is under consideration [21] for installation on the LCLS-II at SLAC [22-24]. In the present work, we consider the operational performance of the nominal cavity formed by four diamond mirrors using both uniform and step-tapered undulator lines for both transmissive and hole out-coupling.

The MINERVA FEL code is used to simulate the interaction in the undulator/FODO line [25,26] and the Optics Propagation Code (OPC) [27-29] is used to propagate the optical field through the resonator. MINERVA/OPC has been validated by comparison with the 10-kW Upgrade experiment at a wavelength of 1.6 μm performed at the Thomas Jefferson National Accelerator Facility [30] where MINERVA/OPC found a maximum output power of 14.52 kW [25] which compared well with the measured value of 14.3 ± 0.7 kW. The capacity to treat Bragg reflections from diamond mirrors has been added to OPC in support of a previous treatment of an XRAFEL [15]. The utility of a tapered undulator in a RAFEL has been studied within the context of an EUV RAFEL design [20].

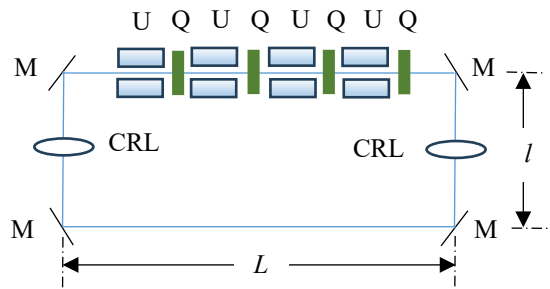

Fig. 1: Schematic of the resonator configuration showing the four diamond mirrors (M1, M2, M3 and M4), the two compact refractive lenses (CRL1 and CRL2), and the undulator/FODO lattice where the undulators (U) and quadrupoles (Q) are indicated.

The resonator under consideration is illustrated in Fig. 1 and consists of four diamond Bragg mirrors located in a rectangular configuration with the mirrors (denoted M1, M2, M3 and M4) angled at 45° with respect to the resonator corners. We denote the length of the long sides as $L$ while that of the short sides as $l$. This corresponds to the resonator under study for installation on the LCLS-II at SLAC [21]. Focusing of the optical field is accomplished by placing two compound refractive lenses (CRL in the figure) in the middle of the short sides of the resonator. The undulator/FODO lattice is composed of a sequence of undulator (planar) segments with strong focusing quadrupoles placed in the center of the drift spaces separating the undulators and is placed along the leg separating mirrors M1 and M4. This is illustrated schematically in the figure where the undulators are designated by U and the quadrupoles by Q. The electron beam is considered to propagate in the direction of mirror

Corresponding author: hfeund@umd.edu

M1 which is designated as the output mirror. Cavity tuning is accomplished by adjusting the length of the long sides of the cavity.

For the sake of the study discussed herein, we assume that the cavity dimensions are $L = 149.0$ m and $l = 1.0$ m and that these dimensions define the zero-detuning length $L_0 = 2(L + l)$ and the synchronous electron beam repetition rate $f_{rep} = c/L_0 \approx 1.0$ MHz.

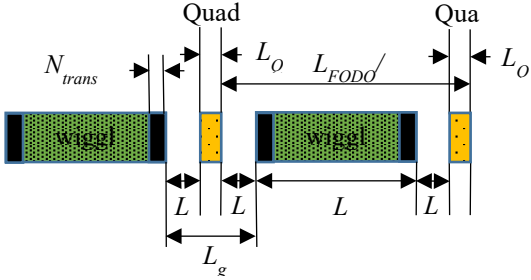

Fig. 2: Illustration of one FODO cell of the undulator line.

| **Undulators** (planar) | HXR |
|---|---|
| Period, $\lambda_w$ | 2.6 cm |
| Length, $L_w$ | 3.432 m |
| On-Axis Field, $B_w$ | 6.83 kG |
| $K$ | 1.66 |
| Entry/Exit transitions | 1 period |
| **Quadrupoles** | |
| Field Gradient | 2.6366 kG/cm |
| Length, $L_Q$ | 13.0 cm |
| FODO Cell Length, $L_{FODO}$ | 7.826 m |
| **Separation Distances** | |
| Drift Length, $L_d$ | 23.4 cm |
| Gap Length, $L_{gap}$ | 59.8 cm |

Table 1: Undulator, FODO cell parameters.

A schematic illustration of a single FODO cell is shown in Fig. 2 and we note that the quadrupoles are centered in the drift space between the undulators. The quadrupole field is simulated with a hard edge while electron propagation through the undulators includes one period adiabatic transitions at each end. The undulator and quadrupole parameters are shown in Table 1. The undulator model used corresponds to the HXR (Hard X-Ray) undulators designed for the LCLS-II at SLAC [31]. Given the dimensions of the undulator/FODO lattice, the total length of the interaction region is about 60 m and the undulator/FODO lattice is centered between mirrors M1 and M4.

The electron beam parameters are shown in Table 2. Note that given the electron energy and the undulator period and strength the FEL resonance occurs at a wavelength of approximately 1.26 Å, corresponding to a photon energy of about 9.84 keV.

| Kinetic Energy | 8.0 GeV |
|---|---|
| Current | 800 A |
| Bunch Duration (parabolic) | 50 fs |
| Bunch Length | 15 μm |
| Repetition Rate | 1.0 MHz |
| Bunch Charge | 26.7 pC |
| Normalized Emittance | 0.25 μm |
| rms Energy Spread | 0.015% |
| rms $x$-Dimension | 14.0 μm |
| Twiss $\alpha_x$ | 1.27 |
| rms $y$-Dimension | 17.7 μm |
| Twiss $\alpha_y$ | −0.80 |

Table 2: Electron beam parameters.

Beam propagation is relatively insensitive to the undulator parameters in the strong focusing FODO lattice, and a plot of the average evolution of the beam envelopes in the $x$- and $y$-directions through the undulator/FODO lattice as determined in simulation is shown in Fig. 3 where the average beam size is seen to be about 15.4 μm. Since the optical mode size will be comparable to that of the electron beam this implies that the Rayleigh range of the mode at the undulator exit is about 50 m.

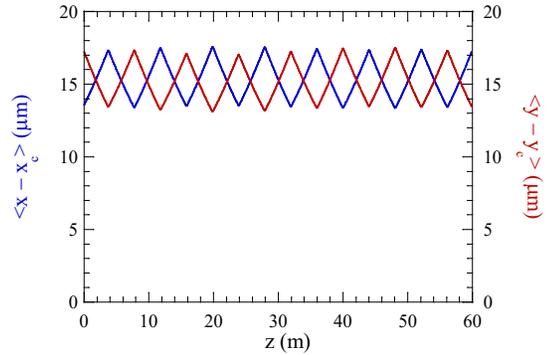

Fig. 3: Evolution of the beam envelope.

The resonator comprises of 4 identical diamond Bragg mirrors with 45° angle of incidence with respect to the (4 0 0) reflecting plane. All mirrors have a thickness of 100 μm, except for the outcouple mirror which has a thickness of 16 μm to allow for extraction of the light from the resonator. The reflection coefficient for the outcouple mirror, with a plane wave incident at 45 degrees, is shown in Fig. 4 where the calculated value of the reflection coefficient and the phase shift [32] is plotted versus the photon energy offset with respect to 9.832 keV. The reflection coefficient for the other mirrors is similar in bandwidth, however the curve more closely resembles a top hat. Figure 4 shows that the outcouple mirror has a narrow reflection bandwidth of slightly less than 100 meV. The maximum reflection coefficient is $|r| \approx 0.966$, resulting in a maximum reflectance of approximately 93%. Furthermore, we observe that the phase shift experienced by the optical field upon reflection is energy dependent.

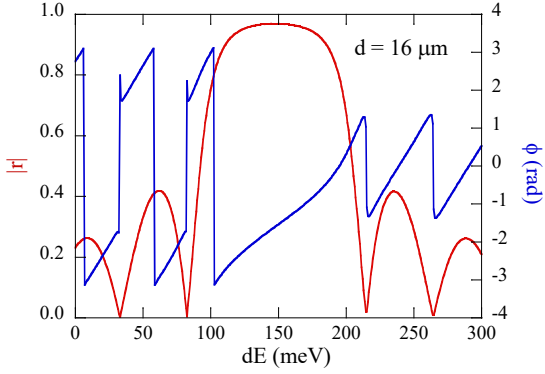

Fig. 4: Reflection coefficient and phase shift versus photon energy offset $dE$ with respect to 9.832 keV for a plane wave incident under 45° with respect to the (4 0 0) crystal plane. The crystal thickness $d$ is 16 µm.

To complete and stabilize the resonator, two compound refractive lenses (see Fig. 1) with a focal length of 106 m are used. To analyse the performance of the resonator, OPC is used to trace an optical pulse, having a photon energy of 9.8319 keV and a parabolic temporal profile with an rms time duration of 20 fs (see Fig. 6), through the cold cavity. Equal mirrors with a thickness $d = 200$ µm are used for this simulation, corresponding to a maximum reflectance of 99.6%. Figure 5 shows the evolution of the rms radius of the optical pulse for one roundtrip through the cavity. We observe that the cold-cavity eigenmode has a minimum rms beam radius of approximately 45 µm at the center of the undulator (and in the opposite arm, half a resonator length away) and has a maximum radius of about 56.5 µm at the location of the lenses.

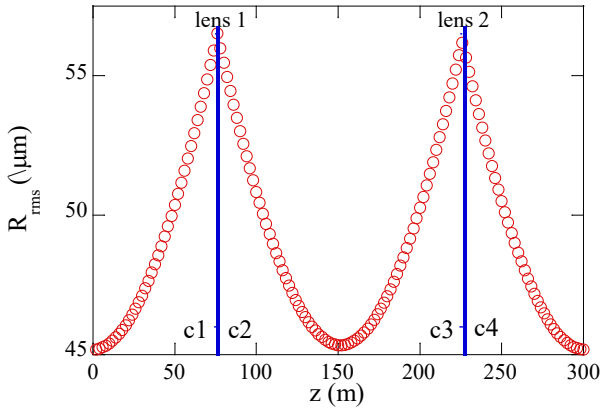

Fig. 5: Root mean square radius of the optical pulse when it travels once through the 300-m long cold. The location of the two compound refractive lenses and the 4 Bragg diamond crystals (C1-C4) are also indicated.

Figure 4 shows that the optical pulse also obtains a phase shift upon reflection, which is similar for the mirrors used for Fig 5. Together with the narrow reflection bandwidth this will affect the temporal profile of the optical pulse. In Fig. 6, we plot the profile of the optical pulse versus position within a comoving frame at the start and end of a single roundtrip, i.e., after 4 reflections from the diamond Bragg mirrors under 45°. The same pulse is used as for Fig. 5. The induced phase change at reflection delays the optical pulse with about 43 fs after one full roundtrip and this will affect the optimum detuning length of the resonator for obtaining maximum pulse energy. Furthermore, some ringing in the optical field can be observed in the tail of the optical pulse. Therefore, care must be taken in setting the width of the simulation window. Finally, the pulse deformation is also affecting the peak power in the pulse. Based on the mirror reflectance we expect to retain 98.4% of the peak power after 4 reflections, however, Fig. 6 shows that this is 97.2%.

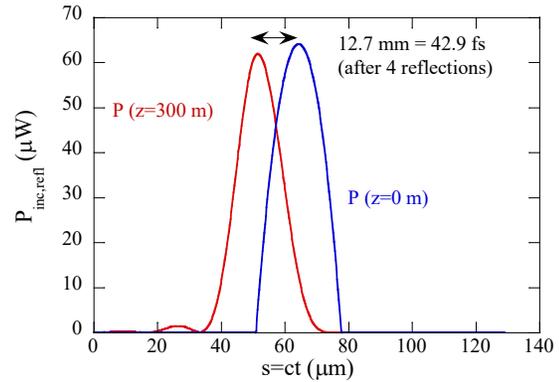

Fig. 6: Temporal profile of the optical pulse at the start and end of one roundtrip through a cold cavity.

Transmissive out-coupling is accomplished by using a thin output mirror (M1) with a thickness of 16 µm (see Fig. 4). In this section we compare the performance of the XRAFEL for both uniform and tapered undulator lines. We assume the undulator line for both the uniform and tapered configurations to consists of 15 segments for a total length of about 59 m. In optimizing the undulator line for the tapered configuration, we must optimize on the start taper segment and the taper slope which we take to be a linear decrease from segment to segment. This does not imply that the tapered performance cannot be improved further by using a nonlinear (e.g., a parabolic) taper but that is a question for a future work. We find optimal performance when the taper starts at the ninth segment followed by a downward slope of 0.07% from one segment to the next.

A plot showing the output energy versus pass for the uniform and tapered undulator lines at zero detuning ($L = L_0$) is shown in Fig. 7. The steady-state is reached after about 6 passes for both configurations. The uniform undulator line saturates at an output energy of about 75.3 µJ with a pass-to-pass fluctuation of 1.1 µJ representing a fluctuation of about 1.5%. This is significantly less than the expected shot-to-shot fluctuation found in a SASE XFEL. The tapered undulator line shows saturation at an output energy of about 141.2 µJ which is enhanced over the uniform undulator configuration by a factor of about 1.9. The pass-to-pass fluctuation for this configuration is about 6.1 µJ for a fluctuation of about 4.3% which is also much less than expected for a SASE XFEL.

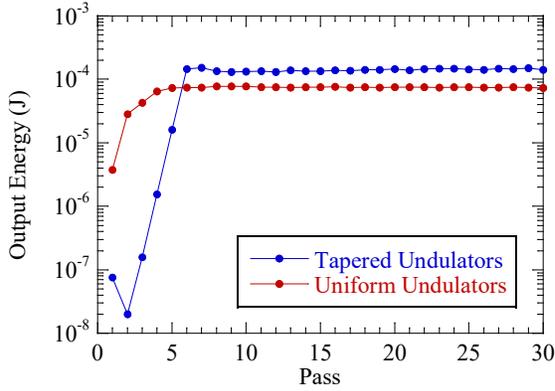

Fig. 7: Output energy versus pass for both uniform and tapered undulator lines for transmissive out-coupling and zero detuning.

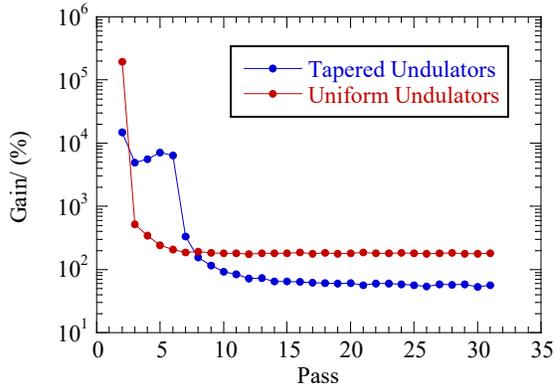

Fig. 8: Gain versus pass for the uniform and tapered undulator lines for zero detuning.

The gain versus pass for the two undulator configurations is shown in Fig. 8, where it is evident that the gain drops rapidly until the steady-state is achieved. The steady state gain is found to be about 182% (60%) for the uniform (tapered) undulator configuration with corresponding losses of about 64% and 36% per pass, respectively. The lower gain in the tapered configuration is due to its shorter uniform section. This will have a significant effect for the hole out-coupling in the tapered undulator configuration as discussed in the next section.

The detuning curve is shown in Fig. 9 where we note that the width of the curve is of the order of 15 μm corresponding to the electron bunch length and that the output energy varies relatively little over this range. We remark that in contrast to low gain/high-Q FEL oscillators, the initial gain is so high that the RAFEL will lase as long as there is some overlap between the incoming electron bunch and the returning optical field. We remark that the "floor" of the curve for $|L - L_0| \geq 8$ μm which corresponds to cavity lengths which are outside the range of overlap between the returning optical pulse and the incoming electron bunch describes single-pass SASE through the undulators.

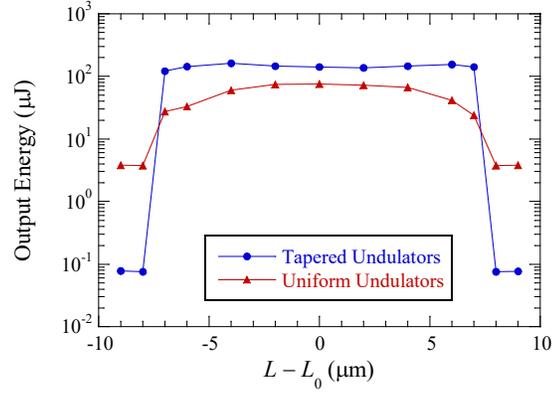

Fig. 9: Detuning curve for transmissive out-coupling.

In simulations of the hole out-coupled configuration, we adopt the same uniform and optimized, tapered undulator lines as used for the transmissive out-coupled simulations. The first order of business is to optimize the radius of the hole, which is placed along the axis of symmetry at the center of the output mirror (M1). Since the hole out-couples the center of the near-Gaussian optical pulse, the reflected/recirculated pulse is hollow. Therefore, the recirculated pulse energy must be large enough that the amplification in the next pass through the undulator line recovers the near-Gaussian pulse shape and reaches saturation (in the steady-state) in a single pass. The optimal hole radius is the maximum possible such that this is achieved [15,16,20].

The variation in the output energy with the hole radius is shown in Fig. 10 for both uniform and tapered undulators at zero detuning. As shown in the figure, the optimum hole radius for the uniform undulator configuration is approximately 100 μm.

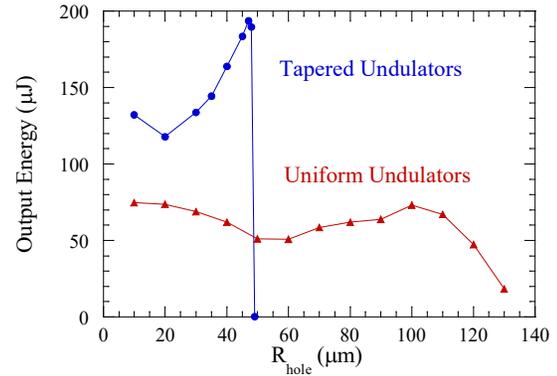

Fig. 10: Output energy versus hole radius for both uniform and tapered undulator lines for hole out-coupling and zero detuning.

However, while the tapered undulator line initially yields a much greater output energy at very small hole radii, the performance crashes rapidly once the hole radius exceeds about 44 μm. The reason for this is that the length of the uniform section of the tapered undulator line is reduced relative to that for the completely uniform case and this results in a reduction in the small signal gain. Thus, once the hole radius exceeds 44 μm, the gain is not high enough to compensate for the losses and the XRAFEL cannot lase. Since (1) fabrication of such small hole radii appears problematical and (2) it seems likely that shot-to-shot jitter in the electron beam may result in a mismatch between the optical mode and the hole even if it were possible. Because of this, we will focus the remainder of this paper on the uniform undulator line.

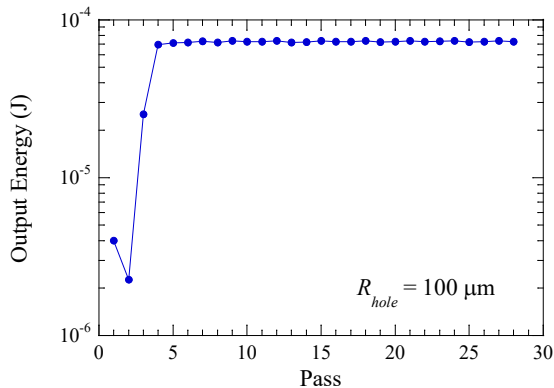

Fig. 11: Output energy versus pass for the uniform a undulator line for hole out-coupling and zero detuning.

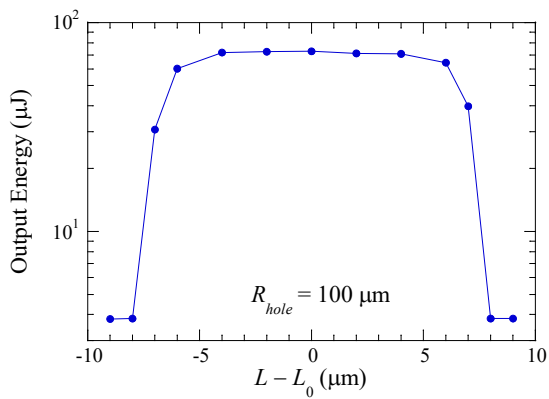

Fig. 12: Detuning curve for hole out-coupling.

The variation in the output energy with pass is shown in Fig. 11 for the optimal hole radius (100 μm) and zero detuning. The average output pulse energy in the steady-state regime is about 72.8 μJ with an rms fluctuation of 0.91 μJ. This represents a pass-to-pass fluctuation of about 1.3% which is much less than would be expected from shot-to-shot fluctuations from a SASE XFEL. Note that this performance is comparable to what was found for the uniform undulator line with transmissive out-coupling.

The detuning curve for hole-out-coupling with uniform undulators is shown in Fig. 12. As in the case of transmissive out-coupling the curve is relatively flat over the range where the returning optical pulse coincides with the incoming electron bunch for $|L - L_0| \leq 5$ μm and drops off rapidly beyond that range. Further, it shows the same "floor" outside this range corresponding to single-pass SASE through the undulator line.

In this paper, we have presented an analysis of an XRAFEL. This is sometimes referred to as a Cavity-Based XFEL (CBXFEL) [31,33-36] although that term can also refer to a low-gain oscillator as well as a RAFEL. The parameters used herein correspond to an experimental configuration under consideration for implementation at the LCLS at SLAC [31]. In this study, we have considered both transmissive and hole out-coupling configurations with, and without, a tapered undulator line. The use of a tapered undulator line was shown to be effective in increasing the efficiency of a RAFEL operating in the EUV regime [20].

The results indicate that the pass-to-pass fluctuations in the XRAFEL for either transmissive or hole out-coupling are less than 5% which is much less than the expected shot-to-shot fluctuations in a SASE XFEL which are typically in the 10 – 15% range. Thus, the output from the XRAFEL will be much more stable than from a SASE XFEL.

The detuning curves for the two types of out-coupling show a relatively flat response with a width of about 14 μm for the configuration under study.

The effect of the taper for transmissive out-coupling is to increase the efficiency by almost 100%. The application of a taper is more problematical for hole out-coupling. We find that the efficiency increases rapidly as the hole radius increases for the parameters under consideration up to about 44 μm but drops off suddenly for larger radii. The reason for this is that the length of the uniform section of the taper is reduced relative to that for the uniform undulator which results in a reduction in the small signal gain, which is not high enough to compensate for the losses, and the XRAFEL cannot lase. However, an alternate resonator design with a higher gain undulator line can make the hole out-coupling/tapered undulator feasible. As such, the XRAFEL constitutes a more stable and efficient source with respect to SASE XFELs.

*Acknowledgments*–This work was supported by the United States Department of Energy under contract DE-SC0024397.